\documentclass[11pt]{article}

\usepackage{natbib}
\setcitestyle{numbers}

\usepackage{graphicx}
\usepackage{amssymb}
\usepackage{latexsym}
\usepackage[left=1.2in,top=1.2in,right=1.2in,bottom=1.2in]{geometry}
\usepackage{amsmath}
\usepackage{booktabs}
\DeclareMathAlphabet{\mathcalligra}{T1}{calligra}{m}{n}

\newcommand{\openr}{\hbox{${\rm I\kern-.2em R}$}}
\newcommand{\openn}{\hbox{${\rm I\kern-.2em N}$}}

    \def\independenT#1#2{\mathrel{\setbox0\hbox{$#1#2$}%
    \copy0\kern-\wd0\mkern4mu\box0}}
\newcommand{\indep}{\rotatebox[origin=c]{90}{$\models$}}

\providecommand{\keywords}[1]
{
	\small	
	\textbf{\textit{Keywords---}} #1
}

\usepackage{abstract}

\title{Targeted Maximum Likelihood Estimation of Community-based Causal Effect of Single Time-Point Community-Level Stochastic Interventionle}

\author{Chi Zhang, Jennifer Ahern, Mark J. van der Laan}
\date{\today}

\begin{document}
\maketitle
 
\begin{abstract}
\noindent
With an increasing need to evaluate the effectiveness of the choice of intervention assigned at the community level in practice, we study the framework for target maximum likelihood estimation and statistical inference for the causal effects of community-level treatments on a community-level outcome defined as the aggregate of the outcomes measured among individuals who are members from the same communities. Current applications in causal inference, especially in the context of hierarchical data structures, have focused on deterministic interventions in which each unit in population receives a fixed value. 
However, positivity violations can easily occur in many cases when certain subgroups in a sample have a (nearly) zero probability of receiving some interventions of interests. Here, we propose a new solution that consider the case in which the treatment mechanism may cause stochastically assigned exposures and the corresponding causal parameter may require a more easily achievable positivity assumption. Then, the target quantity of interest is defined as the mean of a counterfactual community-level outcomes if all communities in the target population receive probabilistically assigned treatments based on a known specified mechanism, which is also called a "stochastic intervention". The causal effect of interest may also be a contrast of the mean of the exposure-specific outcomes under two different stochastic interventions.

Unlike the commonly used parametric regression models such as mixed models, that can easily violate the required statistical assumptions and result in invalid statistical inference, target maximum likelihood estimation allows more realistic data-generative models and provides double-robust, semi-parametric and efficient estimators. Target maximum likelihood estimators (TMLEs) for the causal effect of a community-level static exposure were previously proposed by Balzer \textit{et al} \cite{Balzer_2017}. In this manuscript, we build on this work and present identifiability results and develop two semi-parametric efficient TMLEs for the estimation of the causal effect of the single time-point community-level stochastic intervention whose assignment mechanism can depend on measured and unmeasured environmental factors and its individual-level covariates. The first community-level TMLE is developed under a general hierarchical non-parametric structural equation model, which can incorporate pooled individual-level regressions for estimating the outcome mechanism. The second individual-level TMLE is developed under a restricted hierarchical model in which the additional assumption of "no covariate interference within communities" holds. The proposed TMLEs have several crucial advantages. First, both TMLEs can make use of individual level data in the hierarchical setting, and potentially reduce finite sample bias and improve estimator efficiency. Second, the stochastic intervention framework provides a natural way for defining and estimating casual effects where the exposure variables are continuous or discrete with multiple levels, or even cannot be directly intervened on. Also, the positivity assumption needed for our proposed causal parameters can be weaker than the version of positivity required for other casual parameters.  
\end{abstract}

\keywords{Hierarchical data structure, community-level intervention, stochastic interventions, positivity assumption, nonparametric structural equation model, ensemble learning, targeted maximum likelihood estimation (TMLE).}

\section[Introduction]{Introduction}\label{Introduction1}

\subsection[Motivation]{Motivation}\label{Motivation1}
The literature in fields such as epidemiology, econometrics and social science on the causal impact of community-level intervention, 
is rapidly evolving, both in observational studies and randomized trials. In observation settings, there is a rich literature on assessment of causal effects of families, schools and neighborhoods on child and adolescent development \cite{brooks_aber_duncan_1997, raudenbush_willms_1995}. For instance, the problem addressed by \cite{boyd_wooden_munro_liu_have_2008} is to estimate the impact of community violence exposure on anxiety among children of African American mothers with depression. Similarly, randomized community trials have increased in recent years. As pointed out by \cite{oakes_2004} and \cite{steele_2016}, scientifically speaking, community randomized controlled trials (CRCT) would be a superior strategy estimate the effects of community-level exposures due to self-selection and other difficulties. One example is the MTO study, which estimates the lower-poverty neighborhood effects on crime for female and male youth \cite{kling_ludwig_katz_2005}. Another CRCT example is the ongoing SEARCH study, which estimates the community level interventions for the elimination of HIV in rural communities in East Africa \cite{SEARCH_2013}. Despite recent statistical advances, many of the current applications still rely on estimation techniques such as random effect models (or mixed models) \cite{laird_ware_1982} and generalized estimating equations (GEE) approach \cite{liang_zeger_1986, gardiner_luo_roman_2009}. However, those methods define the causal effect of interest as a coefficient in a most likely misspecified regression model, often resulting in bias and invalid statistical inference in observational settings, and loss of efficiency in randomized community trials. By contrast, the targeted maximum likelihood estimators (TMLE) is constructed based on the efficient influence curve $D^*$, and therefore inherits its double robustness and local efficiency properties. Instead of using $D^*$ directly to construct an efficient estimating equation, TMLE is obtained by constructing a locally least favorable submodel that its score (derivative of the log-likelihood) spans $D^*$ \cite{ bickel_1998, van_der_laan_gruber_2016}.

Deterministic interventions, in which each unit's treatment is set to a fixed value or a value defined by a deterministic function of the covariates, are the main strategy implemented in the current literature for the estimation of causal effects from observational data. One causal assumption needed for parameter identifiability is the positivity assumption. For example, the strong positivity assumption requires that all individuals in the population have a nonzero probability of receiving all levels of the treatment. As argued by \cite{petersen_porter_gruber_wang_van_der_laan_2010}, this strong assumption could be quite unrealistic in many cases. For example, patients with certain characteristics may never receive a particular treatment. On the other hand, a stochastic intervention is one in which each subject receives a probabilistically assigned treatment based on a known specified mechanism. Because the form of the positivity assumption needed for identifiability is model and parameter-specific, stochastic intervention causal parameters are natural candidates if requiring a weaker version of positivity compared to other causal parameters for continuous exposures. Furthermore, a policy intervention will lead to stochastic rather than deterministic interventions if the exposure of interest can only be manipulated indirectly, such as when studying the benefits of vigorous physical activity on a health outcome of interest in the elderly \cite{bembom_van_der_laan_2007}. Because it is unrealistic to enforce every elderly person to have a certain level of physical activity depending on a deterministic rule. To deal with the previous considerations, stochastic interventions could be a more flexible strategy of defining a question of interest and being better supported by the data than deterministic interventions. Thus, using stochastic intervention causal parameters is a good way of estimating causal effects of realistic policies, which could also be naturally used to define and estimate causal effects of continuous treatments or categorical multilevel treatments \cite{munoz_van_der_laan_2011}. 

\subsection[Organization of article]{Organization of article}\label{Organization of article}
The rest of this article is organized as follows. In this article, we apply the roadmap for targeted learning of a causal effect \cite{petersen_van_der_laan_2014}. In Section \ref{Definition of statistical estimation problem} we specify the causal model through a non-parametric structural equation model (NPSEM), allowing us to define the community-level causal effect of interest for arbitrary community-level stochastic interventions as a parameter of the NPSEM, define the corresponding observed data structure, and establish the identifiability of the causal parameter from the observed data generating distribution. We allow for general types of single time-point interventions, including static, dynamic and stochastic interventions. In other words, there is no further restrictions on the intervention distributions, which could be either degenerate (for deterministic interventions) or non-degenerate (for stochastic interventions). Next, Section \ref{Estimation and inference under the general hierarchical causal model} and \ref{Estimation and inference under the restricted hierarchical model with no covariate interference} introduce two different TMLEs of the counterfactual mean outcome across communities under a community level intervention that are based on community-level and individual-level analysis, respectively. Both TMLEs can make use of individual level data in the hierarchical setting. The first community-level TMLE is developed under a general hierarchical causal model and can incorporate some working models about the dependence structure in a community. In other words, the Super Learner library of candidate estimators for the outcome regression can be expanded to include pooled individual-level regressions based on the working model. The first TMLE also includes the case of observing one individual per community unit as a special case. The second individual-level TMLE is developed under a more restricted hierarchical model in which the additional assumption of dependence holds.

\section[Definition of statistical estimation problem]{Definition of statistical estimation problem}\label{Definition of statistical estimation problem}

\subsection[General hierarchical casual model]{General hierarchical casual model}\label{General hierarchical casual model}

Throughout this chapter, we use the bold font capital letters to denote random vectors and matrices. In studies of community-level interventions, we begin with a simple scenario that involves randomly selecting J independent communities from some target population of communities, sampling individuals from those chosen communities, and measuring baseline covariates and outcomes on each sampled individual at a single time point. Also, the number of chosen individuals within each community is not fixed, so communities are indexed with $j = {1,2, ..., J}$ and individual within the $j^{th}$ community are indexed with $i = {1, ..., N_j}$. 

After selection of the communities and individuals, pre-intervention covariates and a post-intervention outcome are measured on each sampled unit.  Because only some of the pre-intervention covariates have clear individual-level counterpart, the pre-intervention covariates separates into two sets: firstly, let denote $W_{j,i}$ the ($1\times{p}$) vector of $p$ such individual-level baseline characteristics, and so ${\boldsymbol{W}}_j = (W_{j,i}: i=1,...,N_j)$ is an $(N_j{\times}p)$ matrix of individual-level characteristics; secondly let $E_j$ represent the vector of community-level (environmental) baseline characteristics that have no individual-level counterpart and are shared by all community members, including the number of individuals selected within the community (i.e., $N_j \in E_j$). Last, $A_j$ is the exposure level assigned or naturally occurred in community $j$ and ${\boldsymbol{Y}}_j = (Y_{j,i} : i=1,...,N_j)$ is the vector of individual outcomes of interest.

In order to translate the scientific question of interest into a formal causal quantity, we first specify a NPSEM with endogenous variables $X = (E, \boldsymbol{W}, A, \boldsymbol{Y})$ that encodes our knowledge about the causal relationships among those variables and could be applied in both observational setting and randomized trials \cite{pearl_1995, pearl_2009}. 
\begin{align}\label{SCMcohort_I}
U &= (U_E, U_{\boldsymbol{W}}, U_A, U_{\boldsymbol{Y}}) \sim P_U \nonumber \\ 
E &=f_E(U_E) \\ 
\boldsymbol{W} &=f_{\boldsymbol{W}}(E,U_{\boldsymbol{W}}) \nonumber \\ 
A &=f_A(E, {\boldsymbol{W}}, U_A) \nonumber \\ 
{\boldsymbol{Y}} &=f_{\boldsymbol{Y}}(E,\boldsymbol{W},A,U_{\boldsymbol{Y}}). \nonumber
\end{align}

where the $U$ components are exogenous error terms, which are unmeasured and random with an unknown distribution $P_U$.  Given an input $U$, the function $F = \{f_E, f_{\boldsymbol{W}}, f_A, f_{\boldsymbol{Y}}\}$ deterministically assigns a value to each of the endogenous variables. For example, model (\ref{SCMcohort_I}) assumes that each individual's outcome $Y$ is affected by its baseline community-level and individual-level covariates $(E, {\boldsymbol{W}})$ together with its community-level intervention(s) and unobserved factors $(A, U_{\boldsymbol{Y}})$. First, while we might have specification of $f_A$, the structural equations  $f_E, f_{\boldsymbol{W}}, f_{\boldsymbol{Y}}$ do not necessarily restrict the functional form of the causal relationships, which could be nonparametric (entirely unspecific), semiparametric or parametric that incorporates domain knowledge. Second, as summarized by \cite{Balzer_2017}, structural causal model (\ref{SCMcohort_I}) covers a wide range of practical scenarios as it allows for the following types of between-individual dependencies within a community: (i) the individual-level covariates (and outcomes) among members of a community may be correlated as a consequence of shared measured and unmeasured community-level covariates $(E, U_E)$, and of possible correlations between unmeasured individual-level error terms $(U_{\boldsymbol{W}}, U_{\boldsymbol{Y}})$, and (ii) an individual $i$'s outcome $Y_{j,i}$ may influence another $l$'s outcome $Y_{j,l}$ within community $j$, and (iii) an individual's baseline covariates $W_{j,l}$ may influence another outcome $Y_{j,i}$. Actually, we can make an assumption about the third type of between-individual dependence, and so the structural equation $f_{\boldsymbol{Y}}$ will be specified under this assumption. More details will be discussed in section (\ref{The individual-level TMLE}). Third, an important ingredient of this model is to assume that distinct communities are causally independent and identically distributed. The NPSEM defines a collection of distributions $(U,X)$, representing the full data model, where each distribution is determined by $F$ and $P_U$ (i.e., $P_{U,X,0}$ is the true probability distribution of $(U,X)$). We denote the model for $P_{U,X,0}$ with $\mathcal{M^F}$.

\subsection[Counterfactuals and stochastic interventions]{Counterfactuals and stochastic interventions}\label{Counterfactuals and stochastic interventions}
$\mathcal{M^F}$ allows us to define counterfactual random variables as functions of $(U,X)$, corresponding with arbitrary interventions. For example, with a static intervention on $A$, counterfactual $\boldsymbol{Y}_a$ can be defined as $f_{\boldsymbol{Y}}(E, {\boldsymbol{W}}, a, U_{\boldsymbol{Y}})$, replacing the structural equation $f_A$ with the constant $a$ \cite{van_der_laan_rose_2011}. Thus, ${\boldsymbol{Y}}_{j,a} = (Y_{j,i,a} : i=1,...,N_j)$ represents the vector of individual-level outcomes that would have been obtained in community $j$ if all individuals in that community had actually been treated according to the exposure level $a$. More generally, we can replace data generating functions for $A$ that correspond with degenerate choices of distributions for drawing $A$, given $U = u$ and $(E, \boldsymbol{W})$, by user-specified conditional distributions of $A^*$. Such non-degenerate choices of intervention distributions are often referred to as stochastic interventions. 

First, let $g^*$ denote our selection of a stochastic intervention identified by a set of multivariate conditional distributions of $A^*$, given the baseline covariates $(E, \boldsymbol{W})$. For convenience, we represent the stochastic intervention with a structural equation, where $A^* = f_{A^*}(E, \boldsymbol{Y}, U_{A^*})$ in terms of random errors $U_{A^*}$, and so define ${\boldsymbol{Y}}_{g^{*}} = f_{\boldsymbol{Y}}(E, \boldsymbol{W}, A^*, U_{\boldsymbol{Y}})$. Then $\boldsymbol{Y}_{j,g^*} = (Y_{j,i,g^*} : i=1,...,N_j)$ denotes the corresponding vector of individual-level counterfactual outcome for community $j$. Second, let $Y^c$ denote a scalar representing a community-level outcome that is defined as a aggregate of the outcomes measured among individuals who are members within a community, and so $Y^c_{g^{*}}$ is the corresponding community-level counterfactual of interest. One typical choice of $Y_{j,g^*}^c$ is the weighted average response among the $N_j$ individuals sampled from community $j$, i.e. $Y_{j,g^*}^c \equiv \sum_{i=1}^{N_j} \alpha_{j,i}Y_{j,i,g^*}$, for some user-specified set of weights $\alpha$ for which $\sum_{i=1}^{N_j} \alpha_{j,i} = 1$. If the underlying community size $N_j$ differs, a natural choice of $\alpha_{j,i}$ is the reciprocal of the community size (i.e., $\alpha_{j,i} = 1/N_j$). 

\subsection[Target parameter on the NPSEM]{Target parameter on the NPSEM}\label{Target parameter on the NPSEM}

We focus on community-level causal effects where all communities in the target population receive the intervention $g^{*}$, then our causal parameter of interest is given by 
\[ \Psi^F(P_{U,X,0}) 
= \mathbb{E}_{U,X}[Y^c_{g^{*}}] 
= \mathbb{E}_{U,X} \Big\{ \sum\limits_{i=1}^{N} \alpha_{i}Y_{i,g^{*}} \Big\} 
\] 
To simply expression,  we use $\alpha_{i}  = 1/N$ in the remainder of article. We also assume (without loss of generality) that the community-level outcome $Y^c$ is bounded in $[0,1]$. If instead $Y^c \in [a,b]$, the the original outcome will be automatically transformed into $Y^{c'} = \frac{Y^c - a}{b-a}$, and our target parameter is corresponding to $Y^{c'}$. Statistical inference such as the point estimate, limiting distribution and confidence interval for the latter target parameter can be immediately mapped into statistical inference for the original target parameter based on $Y^c$, by simply multiplying by $(b-a)$ \cite{Gruber_van_der_Laan_2010}.

One type of stochastic interventions could be a shifted version of the current treatment mechanism $g_0$, i.e., $P_{g^*}(A = a | E, \boldsymbol{W}) = g_0(a  -  \nu(E, \boldsymbol{W}) | E, \boldsymbol{W})$ given a known shift function $\nu(E, \boldsymbol{W})$. A simple example is a constant shift of $\nu(E, \boldsymbol{W}) = 0.5$. Another more complex type could be stochastic dynamic interventions, in which the interventions can be viewed as random assignments among dynamic rules. A simple example corresponding to the previous shift function is $P_{g^*}(A = a | E, \boldsymbol{W}) = g_0(\text{max}\{a  -  0.5, \text{min}(a)\} | E, \boldsymbol{W})$, indicating that shifted exposure $A^*$ is always bounded by the minimum of the observed exposure $A$.

One might also be interested in the contrasts of the expectation of community-level outcome across the target population of communities under different interventions, i.e., 
\[ \Psi^F(P_{U,X,0}) =  \mathbb{E}_{U,X}(Y^c_{g^{*}_1}) - \mathbb{E}_{U,X}(Y^c_{g^{*}_2}) = \mathbb{E}_{U,X} \Big\{ \frac{1}{N} \sum\limits_{i=1}^{N}Y_{i,g^{*}_1} \Big\} - \mathbb{E}_{U,X} \Big\{ \frac{1}{N} \sum\limits_{i=1}^{N}Y_{i,g^{*}_2} \Big\}\]
where $g^{*}_1$ and $g^{*}_2$ are two different stochastic interventions. 

Finally, additive treatment effect is a special case of average causal effect with two static interventions $g_1^{*}(1 | e, \boldsymbol{w}) = 1$ and $g_2^{*}(0 | e, \boldsymbol{w}) = 1$ for any $e \in E, \boldsymbol{W} \in \boldsymbol{W}$, i.e., 
\[ \mathbb{E}_{U,X}(Y^c(1)) - \mathbb{E}_{U,X}(Y^c(0)) = \mathbb{E}_{U,X}[Y^c_{g^{*}_1(1 | e, \boldsymbol{w}) = 1}] - \mathbb{E}_{U,X}[Y^c_{g^{*}_2(0 | e, \boldsymbol{W}) = 1}] \]

\subsection[Link to observed data]{Link to observed data}\label{Link to observed data}
Consider the study design presented above where for a randomly selected community, the observed data consist of the measured pre-intervention covariates, the intervention assignment, the vector of individual-level outcomes. Formally, one observation on community $j$, is coded as  
\[ O_{j,i} = (E_j, W_{j,i}, A_j, Y_{j,i})\]
which follows the typical time ordering for the variables measured on the $i^{th}$ individuals within the $j^{th}$ community.

Assume the observed data consists of $J$ independent and identically distributed copies of $\mathbf{O}_{j} = (E_j, \boldsymbol{W}_{j}, A_j, \boldsymbol{Y}_{j}) \sim P_0$, where $P_0$ is an unknown underlying probability distribution in a model space $\mathcal{M}^I$. Here $\mathcal{M}^I = \{ P(P_{U,X}): P_{U,X} \in \mathcal{M}^F \}$ denotes the statistical model that is the set of possible distributions for the observed data $O$ and only involves modeling $g_0$ (i.e., specification of $f_A$). The true observed data distribution is thus $P_0  = P(P_{U,X,0})$.

\subsection[Identifiability]{Identifiability}\label{Identifiability}
By defining the causal quantity of interest in terms of stochastic interventions (and target causal parameter as a parameter of the distribution $P_{U,X,0}$) on the NPSEM and providing an explicit link between this model and the observed data, we lay the groundwork for addressing the identifiability through $P_0$. 

In order to express $\Psi^F(P_{U,X,0})$ as a parameter of the distribution $P_0$ of the observed data $O$, we now need to address the identifiability of $\mathbb{E}_{U,X}[Y^c_{g^{*}}]$ by adding two key assumptions on the NPSEM: the randomization assumption so called "no unmeasured confounders" assumption (Assumption 1) and the positivity assumption (Assumption 2). The identifiability assumptions will be briefly reviewed here, for details on identifiability, we refer to see \cite{robins_1986, van_der_laan_2010, van_der_laan_2014, munoz_van_der_laan_2011}.

\textbf{Assumption 1.}
\[ A \indep \boldsymbol{Y}_a | E, \boldsymbol{W} \]
where the counterfactual random variable  $\boldsymbol{Y}_a$ represents a collection of outcomes measured on the individuals from a community if its intervention is set to $A=a$ in causal model (\ref{SCMcohort_I}), replacing the structural equation $f_A$ with the constant $a$.

\textbf{Assumption 2.}
\[ \sup\limits_{a \in \mathcal{A}} \frac{g^{*}(a | E, \boldsymbol{W})}{g(a | E, \boldsymbol{W})} < \infty, 
\hspace{0.2cm} \text{almost everywhere} \]
where $g^{*}(a | E, \boldsymbol{W}) = P_{g^{*}}(A=a|E, \boldsymbol{W})$, and assume $\inf\limits_{a \in \mathcal{A}}{g(a | E, \boldsymbol{W})} > \epsilon$ for some small $\epsilon$.

Informally, \textbf{Assumption 1} restricts the allowed distribution for $P_U$ to ensure that $A$ and $Y$ shares no common causes beyond any measured variables in $X = (E, \boldsymbol{W}, A, \boldsymbol{Y})$. For example, assumption 1 holds if $U_A$ is independent of $U_Y$ , given $E, \boldsymbol{W}$. Then, this randomization assumption implies $A^* \indep \boldsymbol{Y}_a | E, \boldsymbol{W}$. In addition, as $P_{g^{*}}(A=a|E, \boldsymbol{W})$ is specified by users in \textbf{Assumption 2}, a good selection of $g^*$ can be used to estimate the causal parameter of interest, but yet does not generate unstable weighting that causes violations of the positivity assumption. Therefore, this posivitiy assumption is easier to achieve compared to other positivity assumptions that other causal parameters used for continuous interventions.

Under \textbf{Assumption 1} and \textbf{2}, jointly with the consistency assumption (i.e., $A = a$ implies $\boldsymbol{Y}_a = \boldsymbol{Y}$), 
\begin{align*}
& P(\boldsymbol{Y}_{g^{*}} = \boldsymbol{y} | A^* = a, E = e, \boldsymbol{W} = \boldsymbol{w})
= P(\boldsymbol{Y}_a = \boldsymbol{y} | A^* = a,E = e, \boldsymbol{W} = \boldsymbol{w}) \\
&= P(\boldsymbol{Y}_a = \boldsymbol{y} | E = e, \boldsymbol{W} = \boldsymbol{w}) 
= P(\boldsymbol{Y} = \boldsymbol{y} | A = a, E = e, \boldsymbol{W} = \boldsymbol{w})
\end{align*}

So our counterfactual distribution $P(\boldsymbol{Y}_{g^{*}} = \boldsymbol{y})$ can be written as:
\begin{align*}
P(\boldsymbol{Y}_{g^{*}} = \boldsymbol{y}) 
&= \int_{e,\boldsymbol{w}} \int_{a} P(\boldsymbol{Y}_{g^{*}} = \boldsymbol{y} | A^* = a, E = e, \boldsymbol{w} = \boldsymbol{w}) g^{*}(a|e,\boldsymbol{w}) d\mu(a) dP_{E, \boldsymbol{W}}(e, \boldsymbol{w})  \\
& \text{by the law of iterated conditional expectation} \\
&= \int_{e,\boldsymbol{w}} \int_{a} P(\boldsymbol{Y}_{a} = \boldsymbol{y} | E = e, \boldsymbol{W} = \boldsymbol{w}) g^{*}(a|e,\boldsymbol{w}) d\mu_a(a) dP_{E, \boldsymbol{W}}(e, \boldsymbol{w}) \\
& \text{by \textbf{assumption 1} and } A^* \indep \boldsymbol{Y}_a | E, \boldsymbol{W}  \\
&= \int_{e,\boldsymbol{w}} \int_{a} P(\boldsymbol{Y} = \boldsymbol{y} | A=a, E=e, \boldsymbol{W}=\boldsymbol{w}) g^{*}(a|e,\boldsymbol{w}) d\mu_a(a) dP_{E, \boldsymbol{W}}(e, \boldsymbol{w}) \\
& \text{by consistency assumption}
\end{align*}
with respect to some dominating measure $\mu_a(a)$.

Then, $\mathbb{E}_{U,X}[\boldsymbol{Y}_{g^{*}}]$ is identified by the G-computational formula \cite{robins_1986}:
\begin{align*}
\mathbb{E}_{U,X}[\boldsymbol{Y}_{g^{*}}] &= \mathbb{E}_{E, \boldsymbol{W}}[ \mathbb{E}_{g^{*}}[\boldsymbol{Y} | A^* = a, E, \boldsymbol{W}] ] \\
&= \int_{e,\boldsymbol{w}} \int_{a} \mathbb{E}_{g^{*}}(\boldsymbol{Y} | a,e,\boldsymbol{w})g^{*}(a|e,\boldsymbol{w}) d\mu_a(a) dP_{E, \boldsymbol{W}}(e, \boldsymbol{w}) 
\end{align*}

This provides us with a general identifiability result for $\mathbb{E}_{U,X}[Y^c_{g^{*}}]$, the causal effect of the community-level stochastic intervention on any community-level outcome $Y^c$ that is some real valued function of the individual-level outcome $\boldsymbol{Y}$: 
\[ \mathbb{E}_{U,X}[Y_{g^{*}}^c] = \mathbb{E}_{U,X}[\sum\limits_{i=1}^N{\alpha_i}{Y_{g^{*},i}}] = \sum\limits_{i=1}^N{\alpha_i}\mathbb{E}_{E,\boldsymbol{W}}[\mathbb{E}_{g^{*}}[Y_i | A^*, E, \boldsymbol{W}]] \equiv \Psi^I(P_0) = \psi^I_0\]

\subsection[The statistical parameter and model for observed data]{The statistical parameter and model for observed data}\label{The statistical parameter and model for observed data}
If we only assume the randomization assumption in the previous section, then the statistical model $\mathcal{M}^I$ is nonparametric. Based on the result of identifiability, we note that $\Psi^I: \mathcal{M}^I \rightarrow \mathbb{R}$ represents a mapping from a probability distribution of $\mathbf{O}$ into a real number, and $\Psi^I(P_0)$ denotes the target estimand corresponding to the target causal quantity $\mathbb{E}_{U,X}[\boldsymbol{Y}_{g^{*}}]$.

Before defining the statistical parameter, we introduce some additional notation. First, we denote the marginal distribution of the baseline covariates $(E, \boldsymbol{W})$ by $Q_{E, \boldsymbol{W}}$, with a well-defined density $q_{E, \boldsymbol{W}}$, with respect to some dominating measure $\mu_y(y)$. There is no additional assumption of independence for $Q_{E, \boldsymbol{W}}$. Second, let $G$ denote the observed exposure conditional distribution for $A$ that has a conditional density $g(A | E,\boldsymbol{W})$.
Third, we assume that all $Y$ within a community are sampled from the distribution $Q_{\boldsymbol{Y}}$ with density given by $q_{\boldsymbol{Y}} (\boldsymbol{Y} | A, E, \boldsymbol{W})$, conditional on the exposure and the baseline covariates $A, E, \boldsymbol{W}$. Now we introduce the notation $P = P_{\tilde{Q},G}$ for $\tilde{Q} = (Q_{\boldsymbol{Y}}, Q_{E,\boldsymbol{W}})$, and the statistical model becomes $\mathcal{M}^I = \{P_{\tilde{Q},G}: \tilde{Q} \in \tilde{\mathcal{Q}}, G \in \mathcal{G}\}$, where $\tilde{\mathcal{Q}}$ and $\mathcal{G}$ denote the parameter space for $\tilde{Q}$ and $G$, respectively, and $\tilde{\mathcal{Q}}$ here is nonparametric. 

Next, we define $G^*$ as the user-supplied intervention with a new density $g^{*}$, which will replace the observed conditional distribution $G$. So $G^*$ is a conditional distribution that describes how each intervened treatment is produced conditional on the baseline covariate $(E, \boldsymbol{W})$. Given $\tilde{Q}$ and $G^*$, we use $\mathbf{O}^* = (O^*_{j,i} = (E_j, W_{j,i}, A^*_j, Y^*_{j,i}): i=1,...,N_j, j=1,,,.,J)$ to denote a random variable generated under the post-intervention distribution $P_{\tilde{Q},G^*}$. Namely, $P_{\tilde{Q},G^*}$ is the G-computation formula for the post-intervention distribution of observed data $\mathbf{O}$ under stochastic intervention $G^*$ \cite{robins_1986},  and the likelihood for $P_{\tilde{Q},G^*}$ can be factorized as: 
\begin{align}\label{Likelihood 1}
p_{\tilde{Q},G^*}(\mathbf{O}^*) = [\prod\limits_{j=1}^{J} q_{\boldsymbol{Y}}(\boldsymbol{Y}_j^{*} | A_j^*, \boldsymbol{W}_j, E_j)][\prod\limits_{j=1}^{J} g^{*}(A_j^* | E_j, \boldsymbol{W}_j)]q_{E,\boldsymbol{W}}(E,\boldsymbol{W})
\end{align}

Thus our target statistical quantity is now defined as  $\psi_0^I = \Psi^I(P_0) = \mathbb{E}_{\tilde{q}_0, g^*}[Y^c_{g^{*}}]$, where $\Psi^I(P_0)$ is the target estimand of the true distribution of the observed data $P_0 \in \mathcal{M}^I$ (i.e., a mapping from the statistical model $\mathcal{M}^I$ to $\mathbb{R}$). We then define $\bar{Q}(A_j, E_j, \boldsymbol{W}_j) = \int_{y} \boldsymbol{y} q_{\boldsymbol{Y}}(\boldsymbol{y} | A_j, \boldsymbol{W}_j, E_j)d\mu_y(y)$ as the conditional mean evaluated under common-in-$j$ distribution $Q_{\boldsymbol{Y}}$, and so $\bar{Q}^c(A, E, \boldsymbol{W}) \equiv E(Y^c | A,E,\boldsymbol{W})$ as the conditional mean of the community-level outcome. Now we can refer to $Q_0 = (\bar{Q}_0^c, Q_{E,\boldsymbol{W}, 0})$ as the part of the observed data distribution that our target parameter is a function of (i.e., with a slight abuse of notation $\Psi^I(P_0) = \Psi^I(Q_0)$), the parameter $\psi_0^I$ can be written as:
\begin{align}\label{parameter_likelihood}
\psi_0^I = \int_{e \in \mathcal{E}, \boldsymbol{w} \in \mathcal{W}} \int_{a \in \mathcal{A}} \bar{Q}_0^c(a, e, \boldsymbol{w}) g^{*}(a|e, \boldsymbol{w})d\mu_a(a) q_{E, \boldsymbol{W}, 0}(e, \boldsymbol{w})d\mu_{e, w}(e, \boldsymbol{w})
\end{align}
with respect to some dominating measures $\mu_a(a)$ and $\mu_{e,w}(e, \boldsymbol{w})$, where $(\mathcal{A}, \mathcal{E}, \mathcal{W})$ is the common support of $(A, E, \boldsymbol{W})$.

Sometimes researchers might be interested in target quantities defined as the difference or ratio of two stochastic interventions. For example, one might define two target estimands $\mathbb{E}_{\tilde{q}_0, g_1^*}[Y^c_{g_1^{*}}]$ and $\mathbb{E}_{\tilde{q}_0, g_2^*}[Y^c_{g_2^{*}}]$ evaluated under two different interventions $g_1^*$ and $g_2^*$, then defining the target quantity as $\mathbb{E}_{\tilde{q}_0, g_2^*}[Y^c_{g_2^{*}}] - \mathbb{E}_{\tilde{q}_0, g_1^*}[Y^c_{g_1^{*}}]$. Actually a generalization of target quantities can be expressed as Euclidean-value functions of a collection $\{ \mathbb{E}_{\tilde{q}_0, g^*}[Y^c_{g^{*}}]: g^{*} \in \mathcalligra{g}\}$, where $\mathcalligra{g}$ denotes a finite set of possible stochastic interventions.

\section[Estimation and inference under the general hierarchical causal model]{Estimation and inference under the general hierarchical causal model}\label{Estimation and inference under the general hierarchical causal model}

In the previous section, we have defined a statistical model $\mathcal{M}^I$ for the distribution of $\mathbf{O}$,  and a statistical target parameter mapping $\Psi^I)$ for which $\Psi^I(P_{Q_0, G^*})$ only depends on $Q_0$ through a relevant part $Q_0 = Q(P_0)$ of $P_0$. Now we want to estimate $\Psi^I(Q_0)$ via a target maximum likelihood estimator (TMLE) and construct an asymptotically valid confidence interval through the efficient influence curve (EIC). Furthermore, we present a novel method for the estimation of the outcome regression in which incorporates additional knowledge about the data generating mechanism that might be known by design.

As a two-stage procedure, TMLE needs to estimate both the outcome regressions $\bar{Q}_0$ and treatment mechanism $g_0$. Since TMLE solves the EIC estimating equation, its estimator inherits the double robustness property of this EIC and is guaranteed to be consistent (i.e., asymptotically unbiased) if either $\bar{Q}_0$ or $g_0$ is consistently estimated. For example, in a community randomized controlled trial $g_0$ is known to be 0.5 and can be consistently estimated, thus its TMLE will always be consistent. Besides, TMLE is efficient when both are consistently estimated. In other words, when $g_0$ is consistent, a choice of the initial estimator for $\bar{Q}_0$ that is better able to approximate the true value $\bar{Q}_0$ may improve the asymptotic efficiency along with finite sample bias and variance of the TMLE \cite{van_der_laan_rubin_2006}.

\subsection[The efficient influence curve $D^*$]{The efficient influence curve $D^*$}\label{The efficient influence curve $D^*$ community-level}

Before constructing a community-level TMLE of $\Psi^I(P_0)$, we must understand its efficient influence curve. The EIC, evaluated at the true distribution $P_0\in \mathcal{M}$, is given by:
\begin{align*}
    D^I(P_0)(\mathbf{O}) &= \frac{g^{*}}{g_0}(A | E, \boldsymbol{W})(Y^c - \bar{Q}^c_0(A, E, \boldsymbol{W})) \\
    & + \mathbb{E}_{g^{*}}[\bar{Q}_0^c(A,E, \boldsymbol{W}) | E, \boldsymbol{W}] - \Psi^I(\mathbb{P}_{Q,g^{*}})
\end{align*}
where
\begin{align*}
    & \mathbb{E}_{g^{*}}[\bar{Q}_0^c(A,E, \boldsymbol{W}) | E, \boldsymbol{W}] = \int_a \bar{Q}_0^{c*}(a, E, \boldsymbol{W}) g^{*}(a | E, \boldsymbol{W}) d\mu_a(a) \\
    & D_Y^{I}(P_0)(\mathbf{O}) = \frac{g^{*}}{g_0}(A | E, \boldsymbol{W})(Y^c - \bar{Q}^c_0(A, E, \boldsymbol{W})) \\
    & D_{E,\boldsymbol{W}}^{I}(P_0)(\mathbf{O}) = \mathbb{E}_{g^{*}}[\bar{Q}_0^c(A,E, \boldsymbol{W}) | E, \boldsymbol{W}] - \Psi^I(\mathbb{P}_{Q,g^{*}})
\end{align*}
Here $D_Y^{I}(P)$ and $D_{E, \boldsymbol{W}}^{I}(P)$ are defined as the projection of the EIC $D^*(P)$ onto the tangent space of $P_{Y|A, E, \boldsymbol{W}}$ at $P\in \mathcal{M}^I$ and $P_{E,\boldsymbol{W}}$ at $P\in \mathcal{M}^I$, given $P=P_{E, \boldsymbol{W}}P_{A|E, \boldsymbol{W}}P_{Y|A,E, \boldsymbol{W}}$, respectively. Note that the projection of the EIC onto the tangent space of $P_{A | E, \boldsymbol{W}}$ (i.e., the exposure mechanism) is zero.

\subsection[The community-level TMLE]{The community-level TMLE}\label{The community-level TMLE}

The community-level TMLE first obtains an initial estimate $\hat{\bar{Q}}^c(A, E, \boldsymbol{W})$ for the conditional mean of the community-level outcome $\bar{Q}_0^c(A,E,\boldsymbol{W})$, and also an estimate $\hat{g}(A|E,\boldsymbol{W})$ of the community-level density of the conditional treatment distribution $g(A | E, \boldsymbol{W})$. The second targeting step is to create a targeted estimator $\hat{\bar{Q}}^{c*}$ of $\bar{Q}^c_0$ by updating the initial fit $\hat{\bar{Q}}^{c}(A,E,\boldsymbol{W})$ through a parametric fluctuation that exploits the information in the estimated density for the conditional treatment distribution $\hat{g}(A|E,\boldsymbol{W})$. The plug-in community-level TMLE is then computed by the updated estimate $\hat{\bar{Q}}^{c*}(A,E,\boldsymbol{W})$ and the empirical distribution of $(E, \boldsymbol{W})$. In this subsection, we describe the community-level TMLE algorithm for estimating the community-based effect under community-level stochastic interventions. For further discussion, please see \cite{munoz_van_der_laan_2011, tmlenet_R, Balzer_2017}.


\subsubsection[Estimation of exposure mechanisms $g_0$ and $g^*_0$]{Estimation of exposure mechanisms $g_0$ and $g^*_0$}\label{Estimation of exposure mechanisms $g_0$ and $g^*_0$ community-level}

A data-adaptive estimator of a conditional density that can be used to estimate the exposure mechanism is proposed by D\'{a}az and van der Laan \cite{munoz_van_der_laan_2011_2}. Here, we build on this work and present how to use the histogram-like estimator to estimate the community-level multivariate exposure mechanism $g_0(A | E, \boldsymbol{W})$. First let's define $g_0(a | E, \boldsymbol{W}) \equiv P_0(A=a | E, \boldsymbol{W})$, where the exposures and baseline covariates $(A, E, \boldsymbol{W})= ((A_j, E_j, \boldsymbol{W}_j) : j = 1,\dots, J)$ denote the random variables drawn jointly from the distribution $S_0(A, E, \boldsymbol{W})$ with the density $s_0(a, e, \boldsymbol{w}) \equiv g_0(a|e,\boldsymbol{w} )q_{E,\boldsymbol{W},0}(e,\boldsymbol{w})$. Here $q_{E,\boldsymbol{W},0}(e,\boldsymbol{w})$ denotes the marginal density of the baseline covariates $(E, \boldsymbol{W})$, and communities are indexed with $j=1,\dots,J$. Then, let's denote $g_0^*(a^* | E, \boldsymbol{W}) \equiv P_{g_0^*}(A=a | E, \boldsymbol{W})$. The fitting algorithm for the non-parametric estimator $g_0^*(A^* | E, \boldsymbol{W})$ is equivalent, except that now the exposures and baseline covariates $(A^*, E, \boldsymbol{W})= ((A^*_j, E_j, \boldsymbol{W}_j) : j = 1, .., J)$ are randomly drawn from $S^*_0(A, E, \boldsymbol{W})$ with the density $s^*_0(a, e, \boldsymbol{w})$ defined as $g^*_0(a|e, \boldsymbol{w})q_{E,\boldsymbol{W},0}(e,\boldsymbol{w})$, where $A^*$ 
is determined by the user-supplied (stochastic) intervention.

Note that $A$ can be multivariate (i.e., $A = (A(m):m=1,\dots,M)$) where $M$ represents the number of treatment variables, and any of its components $A(m)$ can be either binary, categorical or continuous. The joint probability model for $P(A | E, \boldsymbol{W}) \equiv P(A(1),\dots,A(M)| E, \boldsymbol{W})$ can be factorized as a sequence:
\[ P(A(1)| E, \boldsymbol{W}){\times}P(A(2)| A(1), E, \boldsymbol{W}){\times}\dots{\times} P(A(M)| A(1), \dots, A(M-1), E, \boldsymbol{W}) \]

where each of these conditional probability models $P(A(m)| A(1), \dots, A(m-1), E, \boldsymbol{W})$ is fitted separately, depending on the type of the $m$-specific outcome variable $A(m)$. For binary $A(m)$, the conditional probability $P(A(m)| A(1), \dots, A(m-1), E, \boldsymbol{W})$ will be esimtated by a user-specific library of candidate algorithms,  including both parametric estimators and data-adaptive estimators. For continuous (or categorical) $A(m)$, consider a sequence of values $\delta_1, \delta_2, \dots, \delta_{K+1}$ that span the range of $A(m)$ and define $K$ bins and the corresponding $K$ bin indicators, in which case each bin indicator $B_k \equiv [\delta_{k}, \delta_{k+1})$ is used as an binary outcome in a seperate user-specific library of candidate algorithms, with predictors given by $(A(1), \dots, A(m-1), E, \boldsymbol{W})$. That is how the joint probability $P(A | E, \boldsymbol{W})$ is factorized into such an entire tree of binary regression models. 

For simplicity (and without loss of generality), we now suppose $A$ is univariate (i.e., $M=1$) and continuous and  a general template of an fitting algorithm for $P(A | E, \boldsymbol{W})$ is summarized below:

\begin{enumerate}
    \item  Initialization. Consider the usual setting in which we observe $J$ independently and identically distributed copies $\mathbf{o}_{j} = (e_j, \boldsymbol{w}_{j}, a_j, \boldsymbol{y}_{j} : j=1,\dots,J)$ of the random variable $\mathbf{O} = (E, \boldsymbol{W}, A, \boldsymbol{Y})$, where the observed exposure $(a_j:j=1,\dots,J)$ are continuous.
    
    \item Estimation of $P(A=a | E=e, \boldsymbol{W}=\boldsymbol{w})$.
    \begin{enumerate}
        \item  As described above, consider a sequence of $K+1$ values that span the support of $A$ values into $K$ bin intervals $\Delta = (\delta_1,...,\delta_K,\delta_{K+1})$ for a continuous variable $A$. Then any observed data point $a_i$ belongs to one of the $K$ intervals, in other words, for each possible value $a \in A$ (even if this $a$ is not in the observed $(a_j:j=1,\dots,J)$, there always exists a $k \in {1, ...,K}$ such that $a\in [\delta_k, \delta_{k+1})$), and the length (bandwidth) of the interval can be defined as $b_{wk} = \delta_{k+1} - \delta_{k}$.
        
        \item Then let the mapping $S(a)\in \{1,2,\dots,K\}$ denote a unique index of the indicator in $\lambda$ that $a$ falls in, where $S(a)=k$ if $a\in [\delta_k, \delta_{k+1})$, namely $\delta_{S(a)} \leq a < \delta_{S(a)+1}$. Moreover, we use $b_k$ to denote a binary indicator of whether the observed $a$ belongs to bin $k$ (i.e., $b_k\equiv I(S(a)=k)$ for all $k\leq S(a)$). 
        
        \begin{itemize}
            \item This is similar to methods for censored longitudinal data, which treats exposures as censored or missing once the indicator $b_k$ jumps from 0 to 1. 
            \item Since $a$ is a realization of the random variable $A$ for one community, the corresponding random binary indicator of whether $A$ belongs to bin $k$ can be denoted as:
            \[
                B_k = \begin{cases}
                        I(S(A) = k), & \forall k \leq S(A) \\
                        NA, & \forall k > S(A) 
                      \end{cases} 
            \]
        \end{itemize}
        
        \item Then for each $k=1,\dots,K$, a binary nonparametric regression is used to estimate the conditional probability $P(B_k=1|B_{k-1}=0,E,\boldsymbol{W})$, which corresponds to the probability of $B_k$ jumping from 0 to 1, given $B_{k-1}=0$ and the baseline covariates $(E,\boldsymbol{W})$. Here for each $k$, the corresponding nonparametric regression model is fitted only among observations that are uncensored (i.e., still at risk of getting $B_{k}=1$ with $B_{k-1}=0$). Note the above conditional probability 
        \[ P(B_k=1|B_{k-1}=0,E,\boldsymbol{W}) \equiv P(A\in [\delta_{k}, \delta_{k+1}) | A\geq \delta_{k}, E, \boldsymbol{W}) \]
        which is the probability of $A$ belongs to the interval $[\delta_{k},\delta_{k+1})$, conditional on $A$ does not belong to any intervals before $[\delta_{k},\delta_{k+1})$, and $(E,\boldsymbol{W})$.
        
        \item Then the discrete conditional hazard function for each $k$ is defined as a normalization of the conditional probability using the corresponding interval bandwidth $bw_k (\equiv \delta_{k+1} - \delta_{k})$:
        \[ \lambda_k(A, E, \boldsymbol{W}) 
           = \frac{P(B_k = 1|B_{k-1} = 0, E, \boldsymbol{W})}{bw_k}
           = \frac{P(A\in[\delta_{k},\delta_{k+1})|A\geq \delta_{k},E, \boldsymbol{W})}{bw_k} \]
        
        \item Finally, for any given observation $(a, e,\boldsymbol{w})$, we first find out the interval index $k$ to which $a$ belongs (i.e., $k=S(a)\in{1,\dots,K}$). Then the discretized conditional density of $P(A=a | E=e, \boldsymbol{W}=\boldsymbol{w})$ can be factorized by:
        \[ \lambda_k(A, E, \boldsymbol{W})  \times \Big\{ \prod_{t=1}^{k-1}(1 - \lambda_t(A, E, \boldsymbol{W})) \Big\} \]
         which corresponds to the conditional probability of $a$ belongs to the interval $[\delta_{k}, \delta_{k+1})$ and does not belong to any intervals before, given $(E, \boldsymbol{W})$.
    \end{enumerate}
    
    \item The conditional density estimators of $g(A|E,\boldsymbol{W})$ is now proportional to:
    \begin{gather*}
        \prod_{j=1}^{J}P(A_j \in [\delta_{k}, \delta_{k+1}) | E, \boldsymbol{W}) \\ 
        = \prod_{j=1}^{J}\Big[P(A_j\in[\delta_{k},\delta_{k+1})|A_j \geq \delta_{k},E_j, \boldsymbol{W}_j) \times \prod_{t=1}^{k-1}(1 - P(A_j\in[\delta_{t},\delta_{t+1})|A_j \geq \delta_{t},E_j, \boldsymbol{W}_j) \Big] 
    \end{gather*}
    where $P(A\in[\delta_{k},\delta_{k+1})|A \geq \delta_{k}, E, \boldsymbol{W})$ can be estimated by either parametric or data-adaptive algorithms, or the combination of them (i.e., Super Learner). For example, using a main-term only logistic regression: 
    \begin{gather*}
        logit\{ P(A\in[\delta_{k},\delta_{k+1})|A \geq \delta_{k}, E, \boldsymbol{W}) \} \\
        = \sum\limits_{t=1}^k {\alpha_t}I(A\in [\delta_{t-1}, \infty)) + \sum\limits_{s=1}^S{\beta_s}E_s + \sum\limits_{l=1}^p{\gamma_l}\boldsymbol{W}_l
    \end{gather*}
    where we assume that the dimension of $E$ is $S$ and the dimension of $\boldsymbol{W}$ is $p$, and $I(A\in [\delta_{t-1}, \infty))$ indicates if $A$ falls within the interval $[\delta_{t-1}, \infty)$. Alternatively, we can use Super Learner to build a convex combination of the candidate algorithms in the SL library to minimize the cross-validated risk, given a user-specified loss function. 
\end{enumerate}

Note that we need a clever way to determine the bin (interval) cutoffs for a continuous exposure. As proposed by Denby and Mallows \cite{Denby_Mallows_2009}, we can use a histogram-based method that is a compromise between the equal-bin-width histogram and equal-area histogram methods, and the corresponding parameters can be selected by cross validation. For detailed on constructing a histogram-like cross-validated density estimator, we refer to \cite{munoz_van_der_laan_2011_2}.

\subsubsection[Loss function and initial (non-targeted) estimator of $\bar{Q}_0^c$]{Loss function and initial (non-targeted) estimator of $\bar{Q}_0^c$}\label{Loss function and initial (non-targeted) estimator community-level}

As an initial estimator of $\bar{Q}_0^c$, we can simply regress the community-level outcome $Y^c$ onto the exposure and baseline covariates $(A,E,\boldsymbol{W})$. The estimation of $\hat{\bar{Q}}^c$ could be processed by either the usual parametric MLE or loss-based machine learning algorithms based on cross validation, such as loss-based super learning. Given that $Y_j^c$ is bounded continuous or discrete for some known range $Y_j^c \in [a,b], \forall j=1,\dots,J$, the estimation of $\hat{\bar{Q}}^c$ can be based on the following negative Bernoulli log-likelihood loss function: 
\[ -\mathcal{L}^c(\bar{Q}^c)(O) = \sum\limits_{j=1}^J \Big[ Y_j^c\log[\bar{Q}^c(A_j, E_j, \boldsymbol{W}_j)] + (1-Y_j^c)\log[1-\bar{Q}^c(A_j, E_j, \boldsymbol{W}_j)] \Big], \]
or the squared error loss
\[ \mathcal{L}^c(\bar{Q}^c)(O) = \sum\limits_{j=1}^J [Y_j^c - \bar{Q}^c(A_j, E_j, \boldsymbol{W}_j)]^2 \]

For example, for continuous $Y_j^c$, the fitted parameter in a least squares regression can be defined as:
\[ \hat{\beta}^c_{LS} = \arg\min\limits_{\beta}\sum\limits_{j=1}^J [Y_j^c - \bar{Q}_{\beta}^c(A_j, E_j, \boldsymbol{W}_j)]^2 \]

\subsubsection[Loss function and the least favorable fluctuation submodel that spans the efficient influence curve]{Loss function and the least favorable fluctuation submodel that spans the efficient influence curve}\label{Loss function and the least favorable fluctuation submodel that spans the efficient influence curve}

Recall that the targeting step in the TMLE algorithm needs to define a fluctuation parametric submodel for $\hat{\bar{Q}}^c$ and a corresponding user-specified loss function. Given the initial estimator of outcome mechanism $\hat{\bar{Q}}^{c}(A_j, E_j, \boldsymbol{W}_j)$, and the initial estimator of treatment mechanisms $\hat{g}(A_j=a | E_j, \boldsymbol{W}_j)$ and $\hat{g}^{*}(A_j = a | E_j, \boldsymbol{W}_j)$ for each community $j=1,..., J$,  the TMLE algorithm updates the initial estimator $\hat{\bar{Q}}^c$ into $\hat{\bar{Q}}^{c*}$ by
\begin{enumerate}
    \item define a submodel $\hat{\bar{Q}}^c(\epsilon)$ with parameter $\epsilon$ as 
        \[ logit(\hat{\bar{Q}}^{c}(\epsilon)(a, E_j, \boldsymbol{W}_j)) = logit(\hat{\bar{Q}}^{c}(a, E_j, \boldsymbol{W}_j) + \epsilon\hat{H}_j(a,E_j, \boldsymbol{W}_j)), \forall j=1,...,J \]
        where $logit(x) = \log(\frac{x}{1-x})$, and $\hat{H}_j(a,E_j, \boldsymbol{W}_j) = \frac{\hat{g}^{*}(A_j=a | E_j, \boldsymbol{W}_j)}{\hat{g}(A_j=a | E_j, \boldsymbol{W}_j)}$ displays the community-level clever covariate, and the fluctuation parameter $\epsilon$ is obtained by a logistic regression of $Y^c$ on $\hat{H}$ with offset logit($\hat{\bar{Q}}^c$). Note that $\hat{\bar{Q}}^c(\epsilon=0)= \hat{\bar{Q}}^c$ at zero fluctuation.
        
    \item define a community-level loss function such as binary log-likelihood loss function:
        \[ -\mathcal{L}(\hat{\bar{Q}}^{c}(\epsilon))(O) = Y^c\log[\hat{\bar{Q}}^{c}(\epsilon)(A, E, \boldsymbol{W})] +  (1-Y^c)\log [(1-\hat{\bar{Q}}^{c}(\epsilon)(A, E, \boldsymbol{W}))], \]
        and the derivative of the loss function at zero fluctuation has:
        \begin{align*}
            \frac{d}{d\epsilon}\mathcal{L}(\hat{\bar{Q}}^{c}(\epsilon)(O)) |_{\epsilon=0}
            &= \hat{H}(A, E, \boldsymbol{W})(Y^c - \hat{\bar{Q}}^{c}(A, E, \boldsymbol{W})) \\
            &= \frac{\hat{g}^{*}}{\hat{g}}(A | E, \boldsymbol{W})(Y^c - \hat{\bar{Q}}^c(A, E, \boldsymbol{W})) \\
            &= D_Y^I(\hat{Q}, \hat{g})(O)
        \end{align*}
        where $D_Y^I(\hat{Q}, \hat{g})$ is a component of the EIC $D^I$ of $\Psi^I$ at $(\hat{Q}, \hat{g})$.
        
    \item the updated fit $\hat{\bar{Q}}^{c*}$ is defined as $\hat{\bar{Q}}^{c}(\hat{\epsilon}) = logit^{-1}(logit(\hat{\bar{Q}}^{c}) + \hat{\epsilon}\hat{H})$, where $\hat{\epsilon}$ minimizes the empirical loss function above:
        \[ \hat{\epsilon} = \arg\min\limits_{\epsilon} \sum\limits_{j=1}^J  \mathcal{L}(\hat{\bar{Q}}^{c}(\epsilon))(O_j) \]
\end{enumerate}

Another way to achieve the targeting step is to use weighted regression intercept-based TMLE, where $\epsilon$ is obtained by a intercept-only weighted logistic regression of $Y^c$ with offset $logit(\hat{\bar{Q}}^c)(A_j, E_j, \boldsymbol{W}_j)$, predicted weights $\hat{H}_j(A_j, E_j, \boldsymbol{W}_j)$ and no covariates. In summary, this alternative targeting can be implemented by
\begin{enumerate}
    \item define a submodel $\hat{\bar{Q}}^c(\epsilon)$ with parameter $\epsilon$ as 
        \[ logit(\hat{\bar{Q}}^{c}(\epsilon)(a, E_j, \boldsymbol{W}_j)) = logit(\hat{\bar{Q}}^{c}(a, E_j, \boldsymbol{W}_j) + \epsilon, \forall j=1,...,J \].
    \vspace{-0.7cm}
    \item define a weighted (binary log-likelihood) loss function:
        \[ -\mathcal{L}(\hat{\bar{Q}}^{c}(\epsilon))(O) = \Big\{\log[\hat{\bar{Q}}^{c}(\epsilon)(A, E, \boldsymbol{W})^{Y^c}(1-\hat{\bar{Q}}^{c}(\epsilon)(A, E, \boldsymbol{W}))^{1-Y^c}] \Big\} \hat{H}_j(a,E_j, \boldsymbol{W}_j) \] 
    \vspace{-0.7cm}
    \item the updated fit $\hat{\bar{Q}}^{c*} = \hat{\bar{Q}}^{c}(\hat{\epsilon}) = logit^{-1}(logit(\hat{\bar{Q}}^{c}) + \hat{\epsilon})$, where $\hat{\epsilon}$ minimizes the above loss function.
\end{enumerate}

It is worth mentioning that both of the fluctuation methods solves the same empirical EIC estimating equation and thus generate TMLEs with equivalent asymptotic efficiency. However, the latter one, the intercept-based weighted TMLE, is less sensitive to practical positivity violations in finite samples, while obtaining the similar bias reduction in the target parameter \cite{sofrygin_van_der_laan_2016}.

A similar targeting algorithm can be applied to the marginal distribution of $(E,\boldsymbol{W})$. We select a loss function of $Q_{E, \boldsymbol{W}}$ and a parametric working submodel $\hat{Q}_{E, \boldsymbol{W}}(\epsilon)$, so that the derivative of the loss function of $\hat{Q}_{E, \boldsymbol{W}}(\epsilon)$ at zero fluctuation has:
\begin{align*}
    \frac{d}{d\epsilon}\mathcal{L}(\hat{Q}_{E, \boldsymbol{W}}(\epsilon)) |_{\epsilon=0} 
    &= \mathbb{E}_{g^{*}}[\hat{\bar{Q}}^c(A, E, \boldsymbol{W}) | E, \boldsymbol{W}] - \Psi^I(\mathbb{P}_{\hat{Q}, \hat{g}^{*}}) \\
    &= D_{E, \boldsymbol{W}}^I(\hat{Q}, \hat{g})(O)
\end{align*}            
However, this targeting step doesn't generate any update because the empirical distribution $Q_{E, \boldsymbol{W}}$ is non-parametric MLE estimator and has no contribution to the bias for our target parameter \cite{van_der_laan_rose_2011}.

\subsubsection[The community-level TMLE estimator]{The community-level TMLE estimator}\label{The community-level TMLE estimator}

Thus our targeted substitution estimator is computed as the weighted mean of the targeted predictions across the $J$ communities, given the updated estimate $\hat{\bar{Q}}^{c*}$, the estimate of the user-specified stochastic intervention, and the empirical distribution of $(E, \boldsymbol{W})$. One natural choice is the empirical mean defined as follows:
\[  \hat{\Psi}^I(P_{\hat{Q}^{*},\hat{g}^{*}}) = \frac{1}{J} \sum\limits_{j=1}^J \int_{e_j, \boldsymbol{w}_j} \int_{a} \hat{\bar{Q}}^{c*}(a, e_j, \boldsymbol{w}_j) \hat{g}^{*}(a | e_j, \boldsymbol{w}_j) d\mu_a(a) q_{E, \boldsymbol{W}}(e_j, \boldsymbol{w}_j) d\mu_{e,w}(e_j, \boldsymbol{w}_j) \]

\subsubsection[Statistical inference for the community-level TMLE]{Statistical inference for the community-level TMLE}\label{Statistical inference for the community-level TMLE}

By construction, the community-level TMLE estimator $\hat{Q}^{*}$ will solve the EIC equation: \[ 0 = \sum\limits_{j=1}^{J}D^{I}(\hat{Q}^{*},\hat{g}^{*})(O_j)\]
which results in its doubly robust locally efficient property.

In practice, community-level TMLE variance is asymptotically estimated as Var$(\hat{\Psi}^I(\hat{Q}^*))$ $\approx \frac{(\hat{\sigma}^I_J)^2}{J}$, where $(\hat{\sigma}^I_J)^2$ is the sample variance of the estimated influence curve obtained by
\[ (\hat{\sigma}^I_J)^2 = \frac{1}{J} \sum\limits_{j=1}^J \{ D^I(\hat{Q}^*, \hat{g})(O_j)\}^2 \]
where $D^I(\hat{Q}^*, \hat{g})$ is the plug-in estimator of the efficient influence curve of $\Psi^I$ at $P_0$.

This quantity $(\hat{\sigma}^I_J)^2$ can be used to calculate p values and 95\% confidence intervals for different parameters, e.g., $\hat{\Psi}^I(\mathbb{P}_{\hat{Q}^{*},\hat{g}^{*}}) \pm 1.96\frac{\hat{\sigma}^I_J}{\sqrt{J}}$ for the target parameter.

\subsection[Incorporating hierarchical structure for estimating the outcome mechanism]{Incorporating hierarchical structure for estimating the outcome mechanism}\label{Incorporating hierarchical structure for estimating the outcome mechanism}

Based on the previously defined community-level TMLE for the mean of the exposure-specific counterfactual community level outcome, we can still incorporate individual level data rather than simply community wide aggregates of that data. As discussed in section (\ref{Counterfactuals and stochastic interventions}), one typical choice of the community-level counterfactuals of interest is the weighed average response among all individuals sampled from that community, i.e., $Y^c_{j,g^*} = \sum_{i=1}^{N_j}\alpha_{j,i}Y_{j,i,g^*}$. Hence, the conditional mean of the community-level outcome can be rewritten as a weighted average of the individual-level outcomes, $\bar{Q}^c_0(A, E, \boldsymbol{W}) = \mathbb{E}(Y^c | A, E, \boldsymbol{W}) = \sum_{i=1}^{N}\alpha_{i}\mathbb{E}(Y_i | A, E, \boldsymbol{W}) \equiv \sum_{i=1}^{N}\alpha_{i}\mathbb{E}(Y_i | A, E, \boldsymbol{W}, N)$  where the community-specific sample size $N$ is a random variable that is included in the community-level baseline covariates $E$. 

Without changing the underlying structural causal model (1), estimand and efficient influence curve, we may use an individual-level working model to incorporate pooled individual-level outcome regressions as candidates in the Super Learner library for initial estimation of the expected community-level outcome $\bar{Q}^c_0(A, E, \boldsymbol{W})$ given community and individual level covariates, along with community-level exposures. Specially, we propose a working model that assumes that
\begin{align}\label{no_covariate_interference_I}
	\mathbb{E}_0(Y_i | A, E, \boldsymbol{W}) = \mathbb{E}_0(Y_i | A, E, W_i) = \bar{Q}_0(A, E, W_i)
\end{align}
for a common function $\bar{Q}_0$. In practice, this working model suggests that each individual's outcome is drawn from a common distribution that may depend on the individual's baseline covariates, together with the intervention and community-level baseline covariates presented in his or her community, but is not directly influenced by the covariates of others in the same community. 

Furthermore, the strength of the working assumptions could be weakened by encoding the knowledge of the dependence relationship among individuals within communities, namely, defining E to progressively contain a larger subset of any individual-level covariates included in $\boldsymbol{W}$ \cite{Balzer_2017}. For weak covariate interference, the baseline individual-level covariates of other individuals who are connected with individual i could be included into $W_i$. Let $F_i$ denote the subset of individuals whose baseline individual-level covariates affect that individual's outcome $Y_i$, where $i \in F_i$. Now we have a less restricted and more general version of (\ref{no_covariate_interference_I}) as working model:
\begin{align}\label{no_covariate_interference_II}
\mathbb{E}_0(Y_i | A, E, \boldsymbol{W}) = \bar{Q}_0(A, E, (W_l: l \in F_i))
\end{align}
for a common function $\bar{Q}_0$.

We note that this TMLE never claims that the individual-level working model holds, instead, it uses the working model as a means to generate an initial estimator of $\bar{Q}_0^c$. The implementation of the community-level TMLE incorporating hierarchical data is similar to the previous community-level TMLE, except that the estimation of the community-level outcome could also be based on a single pooled individual level regression $Y_{j,i}$ on $(E_j$, $A_j$, $W_{j,i})$ when assuming the aforementioned working model (\ref{no_covariate_interference_I}). As a consequence, the loss functions for the initial estimation of $\bar{Q}^c_0$, can be specified at the individual-level (instead of at the community-level in the previous subsection). For example, we could use a binary log-likelihood loss function:
\[ -\mathcal{L}(\bar{Q}^c)(O) = \sum\limits_{j=1}^J \sum\limits_{i=1}^{N_j} \alpha_{j,i} \Big[ Y_{j,i}\log[\bar{Q}^c(A_{j}, E_j, \boldsymbol{W}_j)] + (1-Y_{j,i})\log[1-\bar{Q}^c(A_j, E_j, \boldsymbol{W}_j)] \Big], \]
or a squared error loss:
\[ \mathcal{L}(\bar{Q}^c)(O) = \sum\limits_{j=1}^J \sum\limits_{i=1}^{N_j} \alpha_{j,i} [Y_{j,i} - \bar{Q}^c(A_j, E_j, \boldsymbol{W}_j)]^2 \]
where $\boldsymbol{\alpha}=(\alpha_{j,i}:j=1,\dots,J, i=1,\dots,N_j)$ is a vector of weights for which $\sum_{i=1}^{N_j}\alpha_{j,i}=1, \forall j$. A common choice of $\alpha_{j,i}$ is $1/N_j$. If the outcome is continuous and we choose the latter loss function, then the fitted parameter $\hat{\beta}_{LS}$ would minimize the above squared error by solving the following first order condition:
\begin{align*}
    \frac{d}{d\beta}\mathcal{L}(\bar{Q}_{\beta}^c)(O) 
    &= \sum\limits_{j=1}^J \sum\limits_{i=1}^{N_j} \alpha_{j,i} \frac{d}{d\beta}[Y_{j,i} - \bar{Q}_{\beta}^c(A_j, E_j, \boldsymbol{W}_j)]^2 \\ 
    &= -2\sum\limits_{j=1}^J \sum\limits_{i=1}^{N_j} \alpha_{j,i} [Y_{j,i} - \bar{Q}_{\beta}^c(A_j, E_j, \boldsymbol{W}_j)]\frac{d}{d\beta}\bar{Q}_{\beta}^c(A_j, E_j, \boldsymbol{W}_j) \\
    &= -2\sum\limits_{j=1}^J \frac{d}{d\beta}\bar{Q}_{\beta}^c(A_j, E_j, \boldsymbol{W}_j) \Big( \sum\limits_{i=1}^{N_j} \alpha_{j,i} [Y_{j,i} - \bar{Q}_{\beta}^c(A_j, E_j, \boldsymbol{W}_j)] \Big) = 0
\end{align*} 
which can easily show that this fitted parameter is identical to the parameter estimated at the community-level (i.e., $\hat{\beta}^c_{LS}$). 

\subsection[Special case where one observation per community]{Special case where one observation per community}\label{Special case where one observation per community}

We will now consider a special case where each community has only one individual (i.e., $N = 1$), and so all individual-level baseline covariates can be treated as environmental factors (i.e., $(E, \boldsymbol{W}) = E$).

\subsubsection[Nonparametric structural equation model]{Nonparametric structural equation model}\label{Nonparametric structural equation model}

Consider a NPSEM with structural equations for endogenous variables $X = (E, A, Y)$,
\begin{align}\label{SCMcohort_III}
E &=f_E(U_E) \\ 
A &=f_A(E, U_A) \nonumber \\ 
Y &=f_{\boldsymbol{Y}}(E, A, U_Y). \nonumber
\end{align}
with endogenous unmeasured sources of random variation $U = (U_E, U_A, U_Y)$.

\subsubsection[Counterfactuals]{Counterfactuals}\label{Counterfactuals}

Let $Y_a = f_Y(E, a, U_Y)$ denote the counterfactual corresponding with setting the treatment $A = a$, thus the community-level counterfactual outcome is the same as the only observation's outcome in community $j$ (i.e., $Y_{j,a}^c \equiv Y_{j,a}$).  

\subsubsection[Observed data]{Observed data}\label{Observed data}

Now the observed data become $O = (E, A, Y)$. We observe $J$ i.i.d observations on $O$.

\subsubsection[Target parameter on NPSEM]{Target parameter on NPSEM}\label{Target parameter on NPSEM}

Consider the following parameter of the distribution of $(U, X)$:
\[ \Psi^F(P_{U,X,0}) = \mathbb{E}_{U,X}[Y^c_{g^{*}}] =\mathbb{E}_{U,X} [Y_{g^{*}}] \] 

\subsubsection[Identifiability Result]{Identifiability Result}\label{Identifiability Result}

\begin{align}
\Psi^F(P_{U,X,0}) &= \mathbb{E}_{U,X}[Y^c_{g^{*}}] = \mathbb{E}_{E}[\mathbb{E}_{g^{*}}[Y | A^*, E]] \equiv \Psi^I(P_0) \nonumber 
\end{align}

\subsubsection[Statistical parameter and efficient influence curve]{Statistical parameter and efficient influence curve}\label{Statistical parameter and efficient influence curve}

\begin{align*}
\Psi^I(P_0) =  \int_{e \in \mathcal{E}} \int_{a \in \mathcal{A}} \bar{Q}_0^c(a, e) g^{*}(a|e)d\mu_a(a) q_{E, 0}(e)d\mu_{e}(e)
\end{align*}
with respect to $\mu_a(a)$ and $\mu_{e}(e)$, where $(\mathcal{A}, \mathcal{E})$ is the common support of $(A, E)$.

The efficient influence curve of the paramter above is:
\begin{align*}
    D^I(P_0)(\mathbf{O}) &= \frac{g^{*}}{g_0}(A | E)(Y^c - \bar{Q}^c_0(A, E)) + \mathbb{E}_{g^{*}}[\bar{Q}_0^c(A,E) | E] - \Psi^I(\mathbb{P}_{Q,g^{*}})
\end{align*}

\subsubsection[Estimation and inference]{Estimation and inference}\label{Estimation and inference}
The TMLE estimator has the same procedure as the previously presented TMLE does, except that here $(E, \boldsymbol{W}) = E$.

\section[Estimation and inference under the restricted hierarchical model with no covariate interference]{Estimation and inference under the restricted hierarchical model with no covariate interference}\label{Estimation and inference under the restricted hierarchical model with no covariate interference}

\subsection[Restricted hierarchical casual Model]{Restricted hierarchical casual model}\label{Restricted hierarchical casual Model}

What if the third type of dependence in model (\ref{SCMcohort_I}) mentioned in section \ref{Estimation and inference under the general hierarchical causal model} is weak or even doesn't exist? This is so called "no covariate interference"  \cite{prague_wang_stephens_tchetgen_degruttola_2016, Balzer_2017}, which describes that each individual's outcome $Y_i$ is sampled from one distribution only depending on the same individual's own baseline covariate $W_i$, the baseline community-level covariates $E$, together with the community-level intervention and that individual's unobserved factors $(A, U_{Y_I})$. Under this working assumption, we have $\mathbb{E}_0(Y_i | A, E, \boldsymbol{W}) = \bar{Q}_0(A, E, W_i)$. Therefore, when background knowledge about $Q_0$ is sufficient to ensure an assumption that working model (\ref{no_covariate_interference_I}) holds, this background changes both the underlying hierarchical causal model and the identifiability results, and so the statistical model, estimand, efficient influence curve, etc. The estimation based on this pooled individual-level regression analysis can leverage the hierarchical data structure and pair the $i$-specific individual-level outcomes and covariates, which may lead to asymptotically more efficient results than a community-level regression analysis.

In this section, we assume such additional knowledge is available and so consider a new hierarchical causal sub-model which restricts the dependence of individuals in a community. The NPSEM that represents the causal relationships among those endogenous variables is now given by:
\begin{align}\label{SCMcohort_II}
E &=f_E(U_E) \nonumber \\ 
\boldsymbol{W} &=f_{\boldsymbol{W}}(E,U_{\boldsymbol{W}}) \\ 
A &=f_A(E, \boldsymbol{W}, U_A) \nonumber \\ 
Y_i &=f_Y(E,W_i,A,U_{Y_i}). \nonumber \\
U_{Y_i} &\indep U_A | E, W_i \nonumber
\end{align}
Here we assume that the conditional distribution of $(W_i, Y_i)$, given $(A,E)$ are common in $i$.

\subsection[Target parameter and the statistical parameter]{Target parameter and the statistical parameter}\label{Target parameter and the statistical parameter}

Let assume that there is a common conditional distribution of $A$ given $(E,W_i)$
across all individuals, i.e., $P(A | E, W_i) \equiv g_{I}(A | E, W)$, where $g_{I}(A | E, W)$ denotes the individual-level stochastic intervention. Recall that we may be interested in $Y_{j,g^*_I}^c \equiv \sum_{i=1}^{N_j} \alpha_{j,i}Y_{j,i,g^*_I}$, with respect to some individual-level stochastic intervention $g^*_I$. We assume that the number of individuals is constant in each community (i.e., $N_j=N, \forall j$). Then our causal parameter of interest is defined by
\[ \Psi^F(P_{U,X,0}) = \mathbb{E}_{U,X}(Y^c_{g^*_I}) = \mathbb{E}_{U,X}(\sum_{i=1}^{N} \alpha_{i}Y_{i,g^*_I}) \]

Note that all of the identifiability results in section (\ref{Identifiability}) can be naturally applied here. Thus, by identifiability, 

\begin{align*}
\Psi^F(P_{U,X,0}) &= \sum_{i=1}^{N}\alpha_{i} \mathbb{E}_{U,X}(Y_{i,g^*_I}) 
\\ 
&= \sum_{i=1}^{N}\alpha_{i}\mathbb{E}_{E, \boldsymbol{W}, 0}\Big\{ \mathbb{E}_{g^*_I}[\bar{Q}_0(A, E, W_i)] \big| E, W_i 
\Big\} \equiv \Psi^{II}(P_{Q,g^*_I}) 
\end{align*}
where $\Psi^{II}: \mathcal{M}^{II} \rightarrow \mathcal{R}$ is the target statistical quantity under the key assumptions of identifiability and working assumption (\ref{no_covariate_interference_I}), and $\mathcal{M}^{II}$ is a sub-model of $\mathcal{M}^I$.

\subsection[The efficient influence curve $D^*$]{The efficient influence curve $D^*$}\label{The efficient influence curve $D^*$ individual-level}

Now, the EIC of $\Psi^{II}$ at $P_0 \in \mathcal{M}^{II}$ is given by:
\begin{align}
    D^{II}(P_0)(\mathbf{O}) &= \sum_{i=1}^{N}\alpha_{i}[\frac{g^*_I}{g_{I,0}}(A | E, W_i)(Y_i - \bar{Q}_0(A, E, W_i)) \nonumber \\ 
    &+ \mathbb{E}_{g^*_I}[\bar{Q}_0(A, E, W_i) | E, W_i] - \Psi^{II}(P_{Q,g^*_I})]  \nonumber 
\end{align}
where
\begin{align*}
    & D_Y^{II}(P_0)(O_i) = \frac{g^{*}_I}{g_{I,0}}(A | E, W_i)(Y_i - \bar{Q}_0(A, E, W_i)) \\
    & D_{E,W}^{II}(P_0)(O_i) = \mathbb{E}_{g^*_I}[\bar{Q}_0(A, E, W_i) | E, W_i] - \Psi^{II}(P_{Q,g^*_I})
\end{align*}
Note that now the EIC is a weighted average of the individual-level EICs.

\subsection[The individual-level TMLE]{The individual-level TMLE}\label{The individual-level TMLE}

\subsubsection[Estimation of exposure mechanisms $g_{I,0}$ and $g_{I,0}^*$]{Estimation of exposure mechanisms $g_{I,0}$ and $g_{I,0}^*$}\label{Estimation of exposure mechanisms $g_{I,0}$ and $g_{I,0}^*$ individual-level}

Here, the individual-level density of the conditional treatment distribution, adjusting for $E$ and the individual specific covariate $W_i$, is defined as 
\begin{align}
g_I(a | e, w_i) &= E_{\boldsymbol{W}}[g_I(a | e, \boldsymbol{W}) | W_i = w_i] = E_{\boldsymbol{W}}[g_I(a | e, \boldsymbol{W}_{-i}, \boldsymbol{W}_{i}) | W_i = w_i]\nonumber \\
&= \int_{\boldsymbol{w}_{-i}} g_I(a | e, \boldsymbol{w}_{-i}, w_i)P(\boldsymbol{W}_{-i} = \boldsymbol{w}_{-i} | W_{i} = w_{i})d\mu(\boldsymbol{w}_{-i}) \nonumber \\
&=  \int_{\boldsymbol{w}_{-i}} g_I(a | e, \boldsymbol{w}_{-i}, w_i)P(\boldsymbol{W}_{-i} = \boldsymbol{w}_{-i})d\mu(\boldsymbol{w}_{-i}) \nonumber
\end{align} 
with respect to some dominating measure $\mu(\boldsymbol{w}_{-i})$, and $\boldsymbol{W}_{-i}$ represents an $((N-1)\times{p})$ matrix of individual-level covariates,  which includes all individuals in the community except that individual $i$. 

Therefore, the estimate of the individual-level stochastic intervention is given by
\begin{align}
\hat{g}_I(a | e, w_i) &= \frac{1}{J}\sum_{j=1}^{J} \int_{\boldsymbol{w}_{j,-i}} \hat{g}_I(a | e_j, \boldsymbol{w}_{j,-i}, w_{j,i})P_n(\boldsymbol{W}_{j,-i} = \boldsymbol{w}_{j,-i})d\mu(\boldsymbol{w}_{j,-i})
\nonumber
\end{align} 
where $\hat{g}_I(a | e_j, \boldsymbol{w}_{j,-i}, w_{j,i})$ can be obtained by the data adaptive methods based histogram-like estimation presented in section (\ref{Estimation of exposure mechanisms $g_0$ and $g^*_0$ community-level}). Besides, the fitting algorithm for $g_{I,0}^*(A^* | E, \boldsymbol{W}_i)$ is equivalent except that $A^*$ is determined by the user-specific stochastic intervention.

\subsubsection[Loss function and initial (non-targeted) estimator of $\bar{Q}_0$]{Loss function and initial (non-targeted) estimator of $\bar{Q}_0$}\label{Loss function and initial (non-targeted) estimator individual-level}

First we assume that the community-level outcome regression is a weighted average of common-in-$i$ individual-level outcome regressions, i,e., $\bar{Q}^c_0 = \sum_{i=1}^N\alpha_i\bar{Q}_0$, where $\bar{Q}_0(A,E,W_i) = \mathbb{E}_0(Y_i|A,E,W_i)$. Therefore, to gain an initial estimator of $\bar{Q}_0$, we can simply regress the individual-level outcome $Y_i$ onto the exposure, the community-level covariates, and the $i$-specific individual-level covariates $(A,E,W_i)$. Without loss of generality, we also assume that $Y_i$ is either bounded continuous or discrete for some known range. Then the estimation can be based on, for example, a squared error loss function:
\[ \mathcal{L}(\bar{Q}_{\beta})(O) = \sum\limits_{j=1}^J \sum\limits_{i=1}^{N_j} \alpha_{j,i} [Y_{j,i} - \bar{Q}_{\beta}(A_j, E_j, W_{j,i})]^2 \]
Here, the fitted parameter $\hat{\beta}_{LS}$ can be solved by minimizing the above squared error function.

\subsubsection[Loss function and the least favorable fluctuation submodel that spans the efficient influence curve]{Loss function and the least favorable fluctuation submodel that spans the efficient influence curve}\label{Loss function and the least favorable fluctuation submodel that spans the efficient influence curve individual-level}

Here, the targeting step again needs to define a fluctuation parametric submodel for $\hat{\bar{Q}}$, the initial estimator of the individual-level outcome regression, and a corresponding pre-specified loss function. Given the initial estimators $\hat{\bar{Q}}(A_j, E_j, W_{j,i})$, $\hat{g}_I(A_j = a | E_j, W_{j,i})$ and $\hat{g}^*_I(A_j = a | E_j, W_{j,i})$, the targeting step will update the individual-level regression estimator $\hat{\bar{Q}}$ into $\hat{\bar{Q}}^*$, and so the community-level regression estimator $\hat{\bar{Q}}^c = \sum_{i=1}^N\alpha_i\hat{\bar{Q}}$ into $\hat{\bar{Q}}^{c*} = \sum_{i=1}^N\alpha_i\hat{\bar{Q}}^*$, by

\begin{enumerate}
    \item define a submodel $\hat{\bar{Q}}(\epsilon)$ with parameter $\epsilon$ as 
        \[ logit(\hat{\bar{Q}}(\epsilon)(a, E_j, W_{j,i}) = logit(\hat{\bar{Q}}(a, E_j, W_{j,i}) + \epsilon\hat{H}_{j,i}(a,E_j, W_{j,i})), \forall j=1,...,J \]
        where $\hat{H}_j(a,E_j, W_{j,i}) = \frac{\hat{g}^{*}(A_j=a | E_j, W_{j,i})}{\hat{g}(A_j=a | E_j, W_{j,i})}$ displays the individual-level clever covariate, and the fluctuation parameter $\epsilon$ is obtained by a pooled logistic regression of the individual-level outcome $Y_i$ on the individual-level covariate $\hat{H}_{i}$ with offset logit($\hat{\bar{Q}}$). Note that $\hat{\bar{Q}}(\epsilon=0) = \hat{\bar{Q}}$ at zero fluctuation.
        
    \item define a loss function for the $i$-specific individual-level outcome regression, such as negative log-likelihood loss function:
        \[ -\mathcal{L}(\hat{\bar{Q}}(\epsilon))(O) = Y_i\log[\hat{\bar{Q}}(\epsilon)(A, E, W_i)] +  (1-Y_i)\log [(1-\hat{\bar{Q}}(\epsilon)(A, E, W_i))], \]
        Then, we use the average of the individual-level loss functions as the loss function for the community-level outcome regression:
        \[ \mathcal{L}^{II}(\hat{\bar{Q}}^c(\epsilon))(O) = \sum_{i=1}^N\alpha_i\mathcal{L}(\hat{\bar{Q}}(\epsilon))(O) \]
        and the derivative of the loss function at zero fluctuation has:
        \begin{align*}
            \frac{d}{d\epsilon}\mathcal{L}^{II}(\hat{\bar{Q}}^{c}(\epsilon)(O)) |_{\epsilon=0}
            &= \sum_{i=1}^N\alpha_i \Big[\frac{\hat{g}_I^{*}}{\hat{g}_I}(A | E, W_i)(Y_i - \hat{\bar{Q}}(A, E, W_i)) \Big] \\
            &= D_Y^{II}(\hat{Q}, \hat{g}_I)(O)
        \end{align*}
         where $D_Y^{II}(\hat{Q}, \hat{g})$ is a component of the EIC $D^{II}$ of $\Psi^{II}$ at $(\hat{Q}, \hat{g}_I)$.
        
    \item the updated fit $\hat{\bar{Q}}^{*}$ is defined as $\hat{\bar{Q}}(\hat{\epsilon}) = logit^{-1}(logit(\hat{\bar{Q}}) + \hat{\epsilon}\hat{H})$, where $\hat{\epsilon}$ minimizes the empirical loss function above:
        \[ \hat{\epsilon} = \arg\min\limits_{\epsilon} \sum\limits_{j=1}^J \sum_{i=1}^{N_j}\alpha_{j,i}  \mathcal{L}(\hat{\bar{Q}}(\epsilon))(O_j) \]
        
    \item the updated community-level regression estimator is $\hat{\bar{Q}}^{c*} = \sum_{i=1}^N\alpha_i\hat{\bar{Q}}^*$.
\end{enumerate}

The weighted regression intercept-based TMLE can be implemented in a similar way as in section (\ref{Loss function and the least favorable fluctuation submodel that spans the efficient influence curve}), except that the targeting step is now based on the individual-level regressions.

Again, applying a similar targeting step to the marginal distribution of $(E,\boldsymbol{W})$ can easily show that the score of $\epsilon$ in the fluctuation model spans the second part of the EIC $D^{II}$:
\begin{align*}
    \frac{d}{d\epsilon}\mathcal{L}(\hat{Q}_{E, \boldsymbol{W}}(\epsilon)) |_{\epsilon=0} 
    &= \sum_{i=1}^N\alpha_i \Big[ \mathbb{E}_{g^{*}}[\hat{\bar{Q}}(A, E, W_i) | E, W_i] - \Psi^I(\mathbb{P}_{\hat{Q}, \hat{g}_I^{*}}) \Big]  \\
    &= D_{E, \boldsymbol{W}}^{II}(\hat{Q}, \hat{g}_I)(O) 
\end{align*}           

\subsubsection[The individual-level TMLE estimator]{The individual-level TMLE estimator}\label{The individual-level TMLE estimator}

The substitution estimator of $\Psi^{II}(P_{Q,g^*_I})$ is defined as follows:
\begin{align}
\Psi^{II}(P_{\hat{Q}^*,\hat{g}^*_I}) = \frac{1}{J}\sum_{j=1}^{J}\sum_{i=1}^{N_j}\alpha_{j,i}\int_{e_j, w_{j,i}} \int_{a_j} \hat{\bar{Q}}^{*}(a, e_j, w_{j,i}) \hat{g}^{*}_I(a | e_j, w_{j,i}) d\mu_{a}(a) q_{E, \boldsymbol{W}}(e_j, \boldsymbol{w}_j) d\mu_{e,w}(e_j, \boldsymbol{w}_j) \nonumber 
\end{align}

\subsubsection[Statistical inference for the individual-level TMLE]{Statistical inference for the individual-level TMLE}\label{Statistical inference for the individual-level TMLE}

Since the individual-level TMLE estimator solves the EIC equation $0 = \sum_{j=1}^JD^{II}(\hat{Q}^*,\hat{g}^*_I)(O_j)$,
the individual-level TMLE variance can be asymptotically estimated as 
\[ Var(\Psi^{II}(\hat{Q}^*)) \approx \frac{(\hat{\sigma}^{II}_J)^2}{J}, \text{ where } (\hat{\sigma}^{II}_J)^2 = \frac{1}{J}\sum\limits_{j=1}^J{D^{II}(\hat{Q}^*,\hat{g}^*_I)(O_j)}^2 \] Then the 95\% confidence interval for $\Psi^{II}(\mathbb{P}_0)$ is $\hat{\Psi}^{II} \pm \frac{\hat{\sigma}^{II}_J}{\sqrt{J}}$.

\bibliographystyle{plain}
\bibliography{tmle_theory}

\end{document}